\documentstyle[12pt]{article}

\newcommand{\EQ}{\begin{equation}}
\newcommand{\EN}{\end{equation}}
\newcommand{\bea}{\begin{eqnarray}}
\newcommand{\ena}{\end{eqnarray}}

\begin{document}
 \def\bq{\begin{quote}}
\def\eq{\end{quote}}
\topmargin -1.2cm
\oddsidemargin 5mm

\renewcommand{\Im}{{\rm Im}\,}
\renewcommand{\thefootnote}{\fnsymbol{footnote}}

\newpage
\begin{titlepage}
\begin{flushright}
IFUM 484/FT\\
November 1994 \\
%hepth@xxx/9205045\\
\end{flushright}
\vspace{1.5cm}\begin{center}
{\bf{{\large CHIRAL INVARIANCE AND LATTICE FERMIONS } \\
{\large WITH MINIMAL DOUBLING}}}\\
\vspace{1cm}
{ M. PERNICI} \footnote{Work supported in part by M.U.R.S.T. and
EEC, Science Project SC1${}^*$-CT92/0789.} \\
\vspace{2mm}
{\em INFN, Sezione di Milano, Via Celoria 16, I-20133 Milano, Italy}\\
\vspace{0.8cm}
\vspace{2cm}
{{\bf{ABSTRACT}}}
\end{center}
\bq
A few years ago some attention has been given to
a fermionic action on the lattice,
with a Wilson-like term which is chirally invariant but
breaks the hypercubic space-time lattice symmetry.
This action describes two Dirac fields in
the continuum limit, provided the coefficient $\lambda$ of the
Wilson-like term satisfies $\lambda > \frac{1}{2}$.

In this letter it is shown that for $\frac{1}{2} < \lambda \leq 1$
the theory is link-reflection positive.
The propagator has the expected real energy poles.
Modulo a phase shift on the fermions, the only relevant terms
which can be added to the action respecting its symmetries
have dimension $4$.

\eq
\vfill
\end{titlepage}
\renewcommand{\thefootnote}{\arabic{footnote}}
\setcounter{footnote}{0}
\newpage

Gauge theories can be studied non-perturbatively using a lattice
approximation \cite{w1} of the euclidean theory.

One of the axioms of euclidean quantum field theory is the
Osterwalder-Schrader reflection positivity condition \cite{o1},
which is needed for continuing the euclidean correlation functions
to Minkowski space.

Lattice QCD with Wilson fermions \cite{w2}
has site-reflection positivity, and a transfer matrix has been
constructed \cite{l1}; it has also link-reflection positivity \cite{o2}.

The disadvantage of using Wilson fermions is that chiral symmetry is
not an exact symmetry. It follows that the partially-conserved-axial-
current approach used in the continuum to explain the existence of
light mesons is not easily implemented on the lattice.
In this respect Kogut-Susskind fermions \cite{s1} give better results,
since there is a residual $U(1)$ axial symmetry in a theory
describing four Dirac fermions in the continuum limit \cite{k1}.

There is a theorem by Nielsen and Ninomiya \cite{n1} stating
that a theory on a space cubic lattice, with a bilinear Hamiltonian which is
local, hermitian, translation invariant, and with bilinear
locally defined conserved charges, has fermions
appearing in pairs, with opposite chirality and the same internal quantum
numbers.
The presence of doublers is related to the fact that the
axial currents are necessarily non-anomalous \cite{k2}.

Karsten has given a space- time lattice version of this theorem \cite{k3}.
In fact, reflection symmetry, hypercubic space-time symmetry,
chiral invariance and locality
impose the presence of a $2^d$ degeneracy of fermions in
$d$ dimensions \cite{p1}.

Giving up the hypercubic symmetry, it is possible to reduce the number of
doublers. In \cite{k3,w3} a model with minimal doubling has been presented;
it contains a Wilson-like term which breaks the hypercubic symmetry
to cubic symmetry.

In this letter it is shown that this model is link-reflection positive
in the range $\frac{1}{2} < \lambda \leq 1 $ of the Wilson-like parameter.
For $\frac{1}{2} < \lambda < 1$ a positive transfer matrix is constructed
explicitly using a double time-slice Hilbert space formalism
\cite{s2}.

The propagator has two real energy poles with the correct continuum
limit; for $\frac{1}{2} < \lambda < 1$
 there are also two complex energy poles,
which decouple in the continuum limit; for $\lambda = 1$ these
extra poles are absent.

Reflection positivity is maintained in presence of gauge fields.

In the continuum limit, the most relevant operator which can be
added to the action respecting its symmetries
is a Lorentz symmetry breaking operator of dimension $3$, which can be
absorbed in the kinetic term with a phase shift on the fermions.

It is argued that, in the continuum limit of lattice QCD,  the fine-tuning
required to recover the Lorentz symmetry in this approach might be easier
to perform than the fine-tuning required to recover the chiral symmetry
in the case of Wilson fermions.

As an example of possible application, a two-flavour QCD
model is mentioned, in which the mirror fermion
 is interpreted as a new flavour; this model
has a $U(1)$ baryon symmetry and an exact $U(1)$ axial symmetry
on the lattice, which is traceless in flavour space, as required
by the cancellation of the axial anomaly.
If the $U(1)$ axial symmetry is spontaneously broken, as it is expected
in the confined phase of QCD, then the
corresponding Goldstone boson has the quantum numbers of a pion.

\vskip 1.2 cm
The naive fermionic action for massless fermions on a space-time
cubic lattice is, in lattice units $a = 1$,
\bea
I_0 = \frac{1}{2} \sum_x \sum_{\mu = 1}^4 {\bar{\psi}}_x \gamma_{\mu}
( \psi_{x + \hat{\mu}} -  \psi_{x - \hat{\mu}} )
\ena
where $x_{\mu}$ is an integer; the gamma matrices are hermitian and
satisfy $\{ \gamma_{\mu}, \gamma_{\nu} \} = 2 \delta_{\mu , \nu}$.
It will be convenient to choose a representation of the gamma
matrices in which $\gamma_1 \gamma_4$ is symmetric; for instance
$\gamma_j = \sigma_2 \otimes \sigma_j$ for $j = 1, 2, 3$
and $\gamma_4 = \sigma_3 \otimes 1$.

The inverse propagator for the naive fermionic action is
\bea
S^{-1}(p) = i \sum_{\mu} \gamma_{\mu} \sin p_{\mu}
\ena
with zeroes for $\sin p_{\mu} = 0$, that is for $p_{\mu} = 0 , \pi$;
it describes $16$ Dirac fields in the continuum limit.
$I_0$ shares this property with any bilinear and
translationally invariant fermionic action,
whose propagator satisfies the following properties \cite{p1}:

i) reflection ($\Theta$ ) symmetry:
$S^{-1}(p_i,p_4) = \gamma_4 S^{-1\dag}(p_i,-p_4) \gamma_4$;

ii) hypercubic space-time lattice symmetry, i.e. invariance
under $\frac{\pi}{2}$ rotations of the coordinate axis,
which together with (i) implies
$S^{-1}(p) = \gamma_{\mu} S^{-1\dag}(R_{\mu} p) \gamma_{\mu}$,
where $R_{\mu}$ is the reflection operator on the $\mu$-th
coordinate, $(R_{\mu}x)_{\nu} = (1 - 2 \delta_{\mu , \nu}) x_{\nu}$;

iii) chiral symmetry:
$S^{-1}(p) = - \gamma_5 S^{-1}(p) \gamma_5$;

iv) locality, in the sense that $S^{-1}(p)$ is continuous
with its first derivatives.

{}From (ii) and (iii) it follows that $S^{-1}(p) = - S^{-1}(-p)$,
which together with periodicity $p_{\mu} \equiv p_{\mu} + 2 \pi$
gives $S^{-1}(\bar{p}) = 0$
for $\bar{p}_{\mu} = 0 , \pi$; therefore a propagator satisfying
these conditions propagates $16$ modes.

Wilson \cite{w2} eliminated this degeneracy introducing a term
which breaks the chiral symmetry,
\bea
I_W = \frac{r}{2} \sum_x \sum_{\mu = 1}^4 \bar{\psi}_x ( 2 \psi_x
- \psi_{x + \hat{\mu}} - \psi_{x - \hat{\mu}} )~~.
\ena
The Wilson action is site-reflection positive for $r = 1$
\cite{l1} and it is link-reflection positive for
$ 0 < r \leq 1 $ \cite{o2} ( for a review see \cite{m1} ).
It describes one massless mode and $15$ massive modes,
which decouple in the continuum limit.
\vskip 0.5 cm

On a cubic  space-time lattice, translation-invariance, locality,
chiral symmetry and CP$\Theta$
(charge-conjugation $\times$ parity $\times$ reflection)
invariance of the action imply the existence of an equal number
of left-handed and of right-handed fermions \cite{k3,p1} .
Therefore under these conditions there is at least a single doubling
of the Dirac modes on the lattice. Then,
in order to have the minimal doubling allowed under these
assumptions, either reflection symmetry or hypercubic invariance
must be dropped.

Reflection symmetry is a necessary ( not sufficient ) condition to
have reflection positivity, which is used to construct a positive
definite transfer matrix, and hence a hermitian Hamiltonian.
While it might be sufficient to have the reflection positivity
condition in the continuum limit, according to the Osterwalder-
Schrader axiom \cite{o1}, to be on the safe side and avoid physical ghosts
 it is better to require reflection positivity on the lattice.
Adding reflection positivity to the above assumptions, it is
necessary to give up hypercubic space-time symmetry, in order
to have less than $15$ doublers.

The Kogut-Susskind fermionic action \cite{s1} is a well-known model
in which the hypercubic space-time lattice symmetry is absent ( there
is however a hypercubic symmetry mixing space-time and flavour indices)
and in which there is an exact $U(1)$ chiral symmetry on the lattice
\cite{k1}.
This action describes four Dirac fermions in the continuum limit.

Karsten \cite{k3} and Wilczek \cite{w3} have given an example
of lattice translation invariant and chirally symmetric fermionic
action which breaks the hypercubic space-time symmetry to
cubic symmetry and which has minimal doubling.  The action is
\bea
I = I_0 + \frac{i \lambda}{2} \sum_x \sum_{\mu \neq 1} \bar{\psi}_x
\gamma_1 ( 2 \psi_x- \psi_{x + \hat{\mu}} - \psi_{x - \hat{\mu}})
\ena
where with respect to the notation in \cite{k3,w3}
the axis $1$ and $4$ are exchanged.
$I_0$ is the naive fermionic action (1).
The inverse propagator is
\bea
S^{-1}(p) = i \sum_{\mu} \sin p_{\mu} \gamma_{\mu}
+ i \lambda \sum_{\mu \neq 1} ( 1 - \cos p_{\mu} ) \gamma_1~~.
\ena
For $\lambda > 1/2$ there are only two propagating
 modes, $p = (0,0,0,0)$ and $p = (\pi,0,0,0)$ \cite{w3} ; in fact the
inverse propagator vanishes provided $\sin p_{\mu} = 0$
for $\mu \neq 1$, which means $p_{\mu} = 0, \pi$ for
$\mu \neq 1$; and provided
$$
\sin p_1 + \lambda \Sigma_{\mu \neq 1} ( 1 - \cos p_{\mu} ) = 0
$$
which cannot be satisfied if $\lambda > 1/2$ and $p_{\mu} = \pi$
for some $\mu \neq 1$.

The hypercubic symmetry is broken to the cubic symmetry in the
directions $x_2, x_3$ and $x_4$, including the axis-inversion symmetry
$\psi_x \rightarrow i \gamma_{\mu} \gamma_5 \psi_{R_{\mu}x}$,
with $\mu \neq 1$.

The action $(4)$ has a discrete symmetry reflecting the fermion
in its mirror fermion:
\bea
\psi_x \rightarrow (-)^{x_1} \psi_{R_1x} ~~~;~~~
\bar{\psi}_x \rightarrow (-)^{x_1} \bar{\psi}_{R_1x}~~.
\ena

The action is link-reflection invariant, that is, invariant under
the antilinear mapping
\bea
\Theta \psi_{\underline{x},t} = \bar{\psi}_{\underline{x},1-t} \gamma_4
{}~~~~~;~~~ \Theta \bar{\psi}_{\underline{x},t} = \gamma_4
\psi_{\underline{x}, 1-t}
\ena
( it is also site-reflection invariant, but not site-reflection
positive ).
It is invariant under $CP$ transformations
$\psi_{\underline{x},t} \rightarrow \gamma_4 C
\bar{\psi}_{-\underline{x},t}^T$ and
$\bar{\psi}_{\underline{x},t} \rightarrow
- \psi_{-\underline{x},t}^T C^{-1} \gamma_4$, where $C$ is the charge
conjugation matrix.
The propagator satisfies the CP-symmetry condition
$$
S^{-1}(p) = \gamma_4 C^{-1} S^{-1 T}(R_4 p) C \gamma_4~~.
$$
Therefore the action is CP$\Theta$-invariant. \\
As in \cite{o2,m1}, define $\psi^{(+)}$ and $\bar{\psi}^{(+)}$ to be the
field variables at times $t \geq 1$ and $\psi^{(-)}$ and $\bar{\psi}^{(-)}$
 those at times $t \leq 0$.
The action decomposes in three parts,
\bea
I = I_{+}[\psi^{(+)}, \bar{\psi}^{(+)}] +
I_{-}[\psi^{(-)}, \bar{\psi}^{(-)}] +
I_c[\psi^{(+)}, \bar{\psi}^{(+)},\psi^{(-)}, \bar{\psi}^{(-)}]
\ena
where
\bea
I_c = \frac{1}{2} \sum_{\underline{x}} [ ( \bar{\psi}_{\underline{x}, 0}
\gamma_4 \psi_{\underline{x}, 1} - \bar{\psi}_{\underline{x}, 1}
\gamma_4 \psi_{\underline{x}, 0} ) - i \lambda
 ( \bar{\psi}_{\underline{x}, 0}
\gamma_1 \psi_{\underline{x}, 1} + \bar{\psi}_{\underline{x}, 1}
\gamma_1 \psi_{\underline{x}, 0} ) ]~~.
\ena
Under the link-reflection symmetry one has
\bea
\Theta \psi_{\underline{x},t}^{(\pm)} =
\bar{\psi}_{\underline{x},1-t}^{(\mp)} \gamma_4 ~~~~; ~~~~
\Theta I_{+} = I_{-} ~~~~; ~~~ \Theta I_c = I_c~~.
\ena
Redefine the Grassmann variables in the following way:
\bea
\xi_{\underline{x},n}^{\dag} = \bar{\psi}_{\underline{x},2n} ~~~;~~~
\xi_{\underline{x},n} = \gamma_4 \psi_{\underline{x},2n+1}~~~;~~~
\eta_{\underline{x},n}^{\dag} = \psi_{\underline{x},2n}^{T} ~~~;~~~
\eta_{\underline{x},n}^{T} = \bar{\psi}_{\underline{x},2n+1} \gamma_4
\ena
which satisfy $\Theta \xi_{\underline{x},n} = \xi_{\underline{x},-n}^{\dag}$
 and $\Theta \eta_{\underline{x},n} = \eta_{\underline{x},-n}^{\dag}$. \\
One has
\bea
I_c = \sum_{\underline{x}} ( \xi_{\underline{x},0}^{\dag} B
\xi_{\underline{x},0} + \eta_{\underline{x},0}^{\dag} B
\eta_{\underline{x},0} )
\ena
where
\bea
B = \frac{1}{2} ( 1 - i \lambda \gamma_1 \gamma_4 ) = B^{\dag} = B^{T}~~.
\ena
$B$ is a positive matrix for $-1 < \lambda < 1$. We will restrict $\lambda$
in the range $\frac{1}{2} < \lambda < 1$ in the following.

For a generic function $F[\psi^{(+)}, \bar{\psi}^{(+)}]$ of the fields at
positive times $t \geq 1$  one has
\bea
\langle (\Theta F ) F \rangle = Z^{-1} \int [ d\bar{\psi}^{(+)} d\psi^{(+)}]
e^{-I_{+}} F[\psi^{(+)}, \bar{\psi}^{(+)}] \times  \nonumber \\
\int [ d\Theta (\bar{\psi}^{(+)}) d \Theta (\psi^{(+)})]
e^{-\Theta I_{+}} F^{\dag}[\Theta \psi^{(+)}, \Theta \bar{\psi}^{(+)}]
\times \nonumber \\
\exp [ - \sum_{\underline{x}} ( \Theta (\xi_{\underline{x},0}) B
\xi_{\underline{x},0} + \Theta (\eta_{\underline{x},0}) B
\eta_{\underline{x},0}) ] \geq 0
\ena
where $e^{-I_{+}} F[\psi^{(+)}, \bar{\psi}^{(+)}]$ depends on
$\xi_{\underline{x},0}$ and $\eta_{\underline{x},0}$, but not on
$\xi_{\underline{x},0}^{\dag}$ and $\eta_{\underline{x},0}^{\dag}$.
Therefore there is link-reflection positivity.

One can construct a positive transfer
matrix translating the fields by two lattice spacings in time,
using a double time-slice Hilbert space formalism \cite{s2} .

Let us consider the Hilbert space which is the Fock space built from
the operator spinor fields $\hat{X}_{\underline{x}}$ and
$\hat{Y}_{\underline{x}}$ satisfying the canonical anticommutation
relations $ \{ \hat{X}_{\underline{x}} , \hat{X}_{\underline{y}}^{\dag} \}
= \delta_{\underline{x}, \underline{y}}$ and
$ \{ \hat{Y}_{\underline{x}} , \hat{Y}_{\underline{y}}^{\dag} \}
= \delta_{\underline{x}, \underline{y}}$
while the other anticommutators vanish.

Define the following transfer matrix operator
\bea
\hat{T} = ( \det B )^{2 V} \exp (-\hat{X}^{\dag} A' \hat{Y}^{\dag T}) ~
\exp ( \hat{X}^{\dag} M \hat{X} + \hat{Y}^{\dag} M \hat{Y} ) ~
\exp ( \hat{Y}^T A' \hat{X} )
\ena
where $V$ is the number of lattice sites at equal time,
\bea
A' = B^{- \frac{1}{2}} A B^{- \frac{1}{2}} ~~~;~~~
e^{M} = B^{- \frac{1}{2}} D B^{- \frac{1}{2}}~~~;~~~
D = \frac{1}{2} ( 1 + i \lambda \gamma_1 \gamma_4 ) = D^{T}
\ena
and $A$ is the following anti-hermitian matrix on equal-time lattice
sites
\bea
A_{\underline{x},\underline{y}} =
\frac{1}{2} \sum_{j=1}^{3} \gamma_j
( \delta_{\underline{y},\underline{x}+\underline{\hat{j}}} -
\delta_{\underline{y},\underline{x}-\underline{\hat{j}}} ) +
i \lambda \gamma_1 \delta_{\underline{y},\underline{x}} +
\frac{i}{2} \lambda \sum_{j=2}^{3} \gamma_1
( 2 \delta_{\underline{y},\underline{x}} -
\delta_{\underline{y},\underline{x}+\underline{\hat{j}}} -
\delta_{\underline{y},\underline{x}-\underline{\hat{j}}} )~~.
\ena

The partition function is defined as
\bea
Z = {\rm Tr}~\hat{T}^{N} =
\int \prod_{\underline{x},n} d\xi_{\underline{x},n}^{\dag}
d\xi_{\underline{x},n} d\eta_{\underline{x},n}^{\dag}
d\eta_{\underline{x},n}~ e^{-I}
\ena
Introduce the Grassmann variables $X_{\underline{x},n}$,
$X_{\underline{x},n}^{\dag}$, $Y_{\underline{x},n}$ and
$Y_{\underline{x},n}^{\dag}$.
Let us use Grassmann coherent states, satisfying
$\hat{X} |X,Y\rangle ~ = X |X,Y\rangle$~,~
$\hat{Y} |X,Y\rangle ~ = Y |X,Y\rangle $~,~\\
$\langle X,Y|\hat{X}^{\dag} =~ \langle X,Y| X^{\dag}$ and
$\langle X,Y|\hat{Y}^{\dag} = \langle X,Y| Y^{\dag}$.

Useful identities are
$$
\langle X,Y| e^{\hat{X}^{\dag} M \hat{X} + \hat{Y}^{\dag} M
\hat{Y}} |X,Y\rangle =
\exp [ X^{\dag} e^{M} X + Y^{\dag} e^{M} Y ]
$$
and
$$
\int dX^{\dag} dX dY^{\dag} dY e^{-(X^{\dag}X + Y^{\dag}Y)}
|X,Y\rangle \langle X,Y| = 1~~.
$$

Using antiperiodic boundary conditions in time, the partition
function is
\bea
Z = \int \prod_{\underline{x},n} dX_{\underline{x},n}^{\dag}
dX_{\underline{x},n} dY_{\underline{x},n}^{\dag}
dY_{\underline{x},n} e^{- \sum_{\underline{x},n}
( X_{\underline{x},n}^{\dag} X_{\underline{x},n} +
Y_{\underline{x},n}^{\dag} Y_{\underline{x},n}) }
\prod_{n} \langle X_{\underline{x},n+1} Y_{\underline{x},n+1} |
\hat{T} | X_{\underline{x},n} Y_{\underline{x},n} \rangle
\ena
giving the action
\bea
I = \sum_{n} [ X_{n}^{\dag} X_{n} + Y_{n}^{\dag} Y_{n} -
X_{n+1}^{\dag} e^M X_{n} - Y_{n+1}^{\dag} e^M Y_{n} +
X_{n+1}^{\dag} A' Y_{n+1}^{\dag T} - Y_{n}^{T} A' X_{n} ]~~.
\ena
For
\bea
X_{\underline{x},n} = B^{\frac{1}{2}} \xi_{\underline{x},n} ~~~;~~~
Y_{\underline{x},n} = B^{\frac{1}{2}} \eta_{\underline{x},n}~~~;~~~
X_{\underline{x},n}^{\dag} = \xi_{\underline{x},n}^{\dag} B^{\frac{1}{2}}
{}~~~;~~
{}~Y_{\underline{x},n}^{\dag} = \eta_{\underline{x},n}^{\dag} B^{\frac{1}{2}}
\ena
the action becomes
\bea
I = \sum_{n} [ \xi_n^{\dag} B \xi_n - \eta_n^T B \eta_n^{\dag T} -
\xi_{n+1}^{\dag} D \xi_n + \eta_n^T D \eta_{n+1}^{\dag T}+
\xi_n^{\dag} A \eta_n^{\dag T} - \eta_n^T A \xi_n ]~~.
\ena
The Jacobian of the transformation cancels the determinant in front
of the transfer matrix operator.

Making the change of variables $(11)$ one obtains the action $(4)$.
Therefore $(15)$ is the transfer matrix for the action $(4)$ with
$\frac{1}{2} < \lambda < 1$.
\vskip 0.5 cm

Given a normal-ordered polynomial
$\hat{\phi} = {\bf :} \! f(\hat{\xi}^{\dag}, \hat{\eta}^{\dag},
\hat{\xi},\hat{\eta} ) {\bf :}$ at $n=0$, that is a polynomial of the field
operators at times $t=0$ and $t=1$, in which $\hat{\xi}^{\dag}$ and
$\hat{\eta}^{\dag}$ go to the left of $\hat{\xi}$ and $\hat{\eta}$,
its time-translate by $2 n$ steps in time is given
by $\hat{\phi}_n = \hat{T}^n \hat{\phi} \hat{T}^{-n}$.
The Schwinger function of a sequence of such operators, for
$n_1 < ..... < n_k$ is given by
\bea
S(\hat{\phi}_{1n_1}...\hat{\phi}_{kn_k}) =
Z^{-1} Tr [ \hat{T}^N \hat{\phi}_{1n_1}...\hat{\phi}_{kn_k} ]
= Z^{-1} Tr [ \hat{T}^{N+n_1} \hat{\phi}_1 \hat{T}^{n_2 - n_1}
\hat{\phi}_2...\hat{\phi}_k \hat{T}^{-n_k} ]  \nonumber \\
= Z^{-1} \int [d\bar{\psi}^{(+)} d\psi^{(+)} d\bar{\psi}^{(-)} d\psi^{(-)}]
\phi_{1n_1} ...\phi_{kn_k} \exp ( - I )~~~~~~~~~~~~~.
\ena

Let us consider the case $\lambda = 1$.
In that case, $B$ is a positive semi-definite matrix,
so that $(14)$ still holds, and the theory is link-reflection positive.
However a transfer matrix cannot be constructed as easily as above,
since $B^{- \frac{1}{2}}$ does not exist; in fact $B$ is a projector
operator for $\lambda = 1$. The situation is somewhat similar to the case
of the Wilson action for $r = 1$.
Using link-reflection positivity, the physical Hilbert space and the
transfer matrix can be constructed in a standard way \cite{o2}.
The physical Hilbert space is defined with the norm
$||F||^2 =~ \langle (\Theta F) F \rangle~ > 0$, where
$F = F[\psi^{(+)}, \bar{\psi}^{(+)}]$ and where the zero norm states have
been factored out. One can associate to $F$ the state $WF$ in the Hilbert
space with the inner product
$( WF, WF ) \equiv \langle (\Theta F) F \rangle~$.
The transfer matrix $T$
is the operator that shifts $F$ by two lattice units in the time
direction. Then the corresponding operator is defined on this Hilbert
space by $( WF, \hat{T}~ WF ) = \langle (\Theta F )(T F) \rangle$.

\vskip 0.4 cm

The physical states correspond to the poles of the propagator in the
energy variable $E = - i p_4$, that is to the real solutions of
\bea
\sinh^2 E = {\sin}^2 p_2 + {\sin}^2 p_3 +
[ \sin p_1 + \lambda ( 3 - \cosh E - \cos p_2 - \cos p_3 ) ]^2~~.
\ena
For $ \frac{1}{2} < \lambda < 1$ there are four roots; the
physical states correspond to the two real energy solutions $ \pm E_1$,
where $E_1 = \ln [ k_1 + \sqrt{k_1^2 - 1} ]$.
There are two complex energy solutions $E = i \pi \pm E_2$, where
$E_2 = \ln [ - k_2 + \sqrt{k_2^2 - 1} ]$
with $k_1 \geq 1$ and $k_2 \leq -1$ given by
\bea
k_{1,2} = \frac{1}{1 - \lambda^2} \{ - \lambda [ \sin p_1 +
\lambda ( 3 - \cos p_2 - \cos p_3 ) ] ~~~~~~~~~~~~~~~~~~~~~~~\nonumber \\
\pm \sqrt{ [\sin p_1 + \lambda ( 3 - \cos p_2 - \cos p_3 ) ]^2 +
( \sin^2 p_2 + \sin^2 p_3 + 1 ) ( 1 - \lambda^2 ) }~ \}~~.
\ena

The two complex poles have a form similar to those of the Wilson
propagator in the range $ 0 < r < 1$ \cite{c1} ;
in particular for
$\underline{p}' = 0$ one gets the same expression  in the two
cases, $E_2 = \ln [ \frac{1 + \lambda}{1 - \lambda} ]$ .
These `time doublers' decouple in the continuum limit.

For $\lambda = 1$, there are only two energy poles, provided
$p_1 \neq - \frac{\pi}{2}$, and $p_2 , p_3 \neq 0$ ; in
the latter case there are no poles. One gets
$E = \pm \ln [ k + \sqrt{k^2 - 1} ]$, where
\bea
k = \frac{\sin^2 p_2 + \sin^2 p_3 +1 +
(3 + \sin p_1 - \cos p_2 - \cos p_3 )^2}{2 (3 + \sin p_1 - \cos p_2 - \cos
p_3)}
\geq 1~~.
\ena
The situation is analogous to the case of the Wilson fermions for $r = 1$.

Taking the continuum limit, one finds that for
$\frac{1}{2} < \lambda \leq 1$ the two real energies tend to the
relativistic values $\pm |\underline{p}'|$, where $|\underline{p}'| << 1$
with  $\underline{p}' = \underline{p}$  or
$\underline{p}' = \underline{p} - (\pi, 0, 0)$,
 which are the positions of the energy poles in the continuum propagator
of the two Dirac modes.

For $\lambda > 1$ there are four real energy poles (and no complex
pole), for any value of $\underline{p}$, instead of the two real
energy poles required to describe a Dirac fermion.
Therefore the dispersion relation is not relativistic.
 This is another way of seeing, without referring to
reflection positivity, that the case $\lambda > 1$ should be excluded.
\vskip 0.4 cm

Let us discuss the continuum limit of this model.

Let $\Psi (x)$ denote the Dirac fermion which is the continuum
limit of the lattice fermion $\psi_x$ with momentum support
near the pole $p = 0$ of the lattice propagator;
the mirror fermion $\Phi (x)$ is the continuum
limit of $(-)^{x_1}i \gamma_1 \gamma_5 \psi_x$ , with $\psi_x$
near the pole $p = (\pi, 0, 0, 0)$.
Under the discrete symmetry $(6)$ the fermion $\Psi$ is transformed
in the mirror fermion $\Phi$, ~
$\Psi (x) \rightarrow i \gamma_1 \gamma_5 \Phi (R_1 x)$.
The fields $\Psi$ and $\Phi$ have opposite chiral charge.

In the continuum limit the action $(4)$ becomes the Dirac action
\bea
I = \int d^4 x \bar{\Psi}(x) \gamma_{\mu} \partial_{\mu} \Psi (x) +
\bar{\Phi} (x) \gamma_{\mu} \partial_{\mu} \Phi (x)
\ena
where the irrelevant higher-derivative terms coming from $(4)$ have
been neglected; therefore in the continuum limit Lorentz invariance
is restored.

The most relevant operator which can be added to the continuum action
$(27)$ and which respects all the previously mentioned
symmetries of the action $(4)$ is
\bea
i \alpha \int d^4 x [ \bar{\Psi}(x) \gamma_1 \Psi (x) -
\bar{\Phi}(x) \gamma_1 \Phi (x) ]
\ena
where $\alpha$ is a real constant. In particular this term is invariant
under the discrete symmetry $(6)$.
No other operator of dimension $3$ respects all required symmetries;
for instance $ \int d^4 x \bar{\Psi}(x) \gamma_1 \gamma_5 \Psi (x)$
breaks the inversion symmetry,
$\int d^4 x [\bar{\Psi}\gamma_5 \Phi + \bar{\Phi} \gamma_5 \Psi ] (x)$
breaks the discrete symmetry $(6)$, while
$ \int d^4 x [ \bar{\Psi} \gamma_5 \Phi -
\bar{\Phi} \gamma_5 \Psi ] (x)$ is the continuum limit of
$\sum_x (-)^{x_1}\bar{\psi}_x \gamma_1 \psi_x$, which
breaks lattice translation invariance.

If the Lorentz symmetry breaking
term $(28)$ is added to the Dirac action $(27)$, it can be
absorbed in the kinetic term by a phase shift
\bea
\Psi (x) \rightarrow \exp [-i\alpha x_1]~ \Psi (x)~~~;~~~
\Phi (x) \rightarrow \exp [+i\alpha x_1]~ \Phi (x)
\ena
recovering Lorentz invariance. Modulo this shift, the only relevant
Lorentz symmetry breaking term which can be added to the action $(4)$
respecting its symmetries is the dimension $4$ operator
$\int d^4 x \bar{\Psi}(x) \gamma_1 \partial_1 \Psi (x) +
\bar{\Phi} (x) \gamma_1 \partial_1 \Phi (x)$.

Few remarks are in order.
\vskip 0.4 cm
i) In \cite{k3,w3} the Wilson-like term is added to the action in the
time direction. In that case one can define a reflection symmetry
in any one of the space directions, but not in the time direction;
since reflection symmetry in time is needed to have the usual
reflection positivity, and a transfer matrix translating fields
in the time direction, in this letter the coordinates $1$ and $4$
have been interchanged with respect to those in \cite{k3,w3}.
A related reason for which $4$ is not the true Euclidean time
direction in \cite{k3,w3} is that rotating $p_4$ to $E = - i p_4$, the
fermionic propagator in \cite{k3,w3} has no real poles for generic
value of $\underline{p}$.
\vskip 0.4 cm
ii) One can half the number of propagating modes introducing a chiral
projector $P_{\pm} = \frac{1}{2} ( 1 \pm \gamma_5 )$.
Defining
$$
\xi_{\underline{x},n}^{\dag} = \bar{\psi}_{\underline{x},2n} P_{-}~~~;~~~
\xi_{\underline{x},n} = \gamma_4 P_{+} \psi_{\underline{x},2n+1}~~~;~~~
\eta_{\underline{x},n}^{\dag} = \psi_{\underline{x},2n}^{T} P_{+}~~~;~~~
\eta_{\underline{x},n}^{T} = \bar{\psi}_{\underline{x},2n+1} \gamma_4 P_{+}
$$
and proceeding with the same Hilbert space construction as above, one gets the
action $(4)$, with $\psi_x$ substituted by $P_{+} \psi_x$; it describes
a right-handed fermion and its mirror fermion, which is left-handed.
Together they give one Dirac fermion.
\vskip 0.4 cm
iii) A mass term, which breaks softly the axial symmetry,
\bea
m \sum_x \bar{\psi}_x \psi_x =
m \sum_{\underline{x},n}
( \xi_{\underline{x},n}^{\dag} \eta_{\underline{x},n}^{\dag T} +
 \eta_{\underline{x},n}^T \xi_{\underline{x},n} ) \simeq
m \int d^4 x [ \bar{\Psi} \Psi + \bar{\Phi} \Phi ] (x)
\ena
can be added, maintaining reflection positivity ( since the mass term
goes into $I_{+} + I_{-}$, not in $I_c$ ) and respecting all remaining
symmetries of the action $(4)$. The energy poles are given by the
same expressions given above, but with $\sin^2 p_2 + \sin^2 p_3$
replaced by $\sin^2 p_2 + \sin^2 p_3 + m^2$. The two real poles
have the correct continuum limit.

\vskip 0.5 cm
iv) It is straightforward to introduce gauge fields, placing as usual
the gauge variables on the links. Quantizing the gauge field
in the gauge $U_{\underline{x},\hat{4}} = 1$ on the links between
$t = 0$ and $t = 1$, the proof of link-reflection
positivity is the same as in $(14)$, since $B$ does not depend on the
gauge variables in this gauge.

To obtain the correct continuum limit, one
must introduce new relevant operators, which have cubic but not
hypercubic symmetry; in absence of fermions, the only such term is
$\sum_{\mu \neq 1}F_{\mu ,1}^2$ \cite{w3}.
After the phase shift $(29)$, all the hypercubic-breaking
relevant operators have dimension $4$.
If one computes correlation functions of composite operators
which are invariant under the phase transformation $(29)$, as for instance
$\langle \bar{\Psi}(x) \Gamma \Psi (x')
\bar{\Psi}(y) \Gamma \Psi (y') \rangle $, with $x_1 = x_1'$ and $y_1 = y_1'$,
the transformation $(29)$ is not observable, and the
renormalized quantities do not depend on it. It follows that in the
renormalization of these correlation functions only
hypercubic-symmetry-breaking operators of dimension $4$ must
be added; the corresponding counterterms are at most logarithmically divergent.
The situation is similar to the case of Wilson fermions; in both cases
there is a dimension $5$ operator which breaks a symmetry, the
hypercubic symmetry in the present case, the chiral symmetry in
the case of Wilson fermions.
In the latter case, the
most relevant axial-symmetry breaking operator which can be introduced
by radiative corrections is $\bar{\Psi} \Psi$,
which has dimension $3$; the mass counterterm is linearly
divergent, so that the fine-tuning needed to recover the chiral
symmetry is more difficult to perform than the fine-tuning required
in the presently discussed case.
These issues will be studied further.
\vskip 0.4 cm

v) As a possible application, consider the action $(4)$ with
fermions in the irreducible representation $\underline{N}$
of the colour group $SU(N)$. In the continuum limit
it describes two `flavours', that is the two continuum Dirac fields
described above.
On the lattice there is a $U(1)$ baryon symmetry and a $U(1)$ axial
symmetry, which is traceless in flavour space, since a fermion and
its mirror fermion have opposite chiral charges.
 Following \cite{k1,k4} it can be expected that a strong-coupling
analysis would show, in $\frac{1}{N}$ or $\frac{1}{d}$ expansion,
 that the axial current is broken spontaneously in
a dynamical way. The corresponding Goldstone boson is a flavour non-singlet
pseudo-scalar.
In the usual PCAC interpretation, adding the mass term $(30)$ to the
model, the Goldstone boson becomes a pseudo-Goldstone boson, which
can be interpreted as a low-mass pion.

\newpage
I would like to thank C. Destri and G. Marchesini for discussions.

\end{document}